\documentclass[twocolumn,showpacs,preprintnumbers,amsmath,amssymb,showkeys]{revtex4}

\usepackage{graphicx}
\usepackage{dcolumn}
\usepackage{bm}

\usepackage{amsfonts}
\usepackage{amssymb,amsmath}

\newcommand{\be}{\begin{equation}}
\newcommand{\ee}{\end{equation}}
\newcommand{\bdis}{\begin{displaymath}}
\newcommand{\edis}{\end{displaymath}}

\newcommand{\HBM}{\hat{H}_M}

\begin{document}

\title{Coleman - de Luccia instanton of the second order in a brane world}

\author{Michal Demetrian}\email{demetrian@fmph.uniba.sk}
\affiliation{Comenius University \\
Mlynska Dolina M105, 842 48, Bratislava, Slovak Republic }

\date{\today}

\begin{abstract}
The second order Coleman - de Luccia instanton and its action in the Randall - Sundrum type II model
are investigated and the
comparison with the results in Einstein's general relativity is done in the present paper.
\end{abstract}

\maketitle

\section{Introduction}

The instability of the false vacuum of a scalar field interacting with gravity can result in the formation of rapidly
expanding bubbles containing the scalar field (inflaton) on the true vacuum side of the potential barrier,
\cite{cdl}. This process, described by the Coleman - de Luccia (CdL) instanton,
takes place (the CdL instanton exists) under some conditions on the inflaton potential $V$,
\cite{js}, \cite{tanaka}, \cite{vj1}. There is also another way in which the false vacuum can decay. This one is
mediated by the Hawking - Moss (HM) instanton which exists for any non-negative potential $V$ obeying two
non-degenerate minima separated by a finite barrier.
The false vacuum decay via the HM instanton
describes the inflaton that jumps at the top of the potential barrier within the horizon-size domain, \cite{hm},
\cite{linde}, and afterwards rolls down-hill to the true vacuum. \\
The semiclassical theory of the false vacuum decay by the instantons can be formulated also in the brane world scenarios,
for further details se \cite{diaz}, \cite{cahesa} and \cite{jabv}.
The equations for the CdL instantons in the Randall-Sundrum type II scenario have the form \cite{cahesa}
\begin{eqnarray} \label{eebrane}
& &
a''=-C\left\{ (\Phi')^2+V+\frac{1}{8\sigma}\left[(5(\Phi')^2+2V) \right. \right. \nonumber \\
& &
\left. \left. (-(\Phi')^2+2V)\right]\frac{}{}\right\}a,\nonumber \\
& &
\Phi''+3\frac{a'}{a}\Phi'-\partial_\Phi V=0,
\end{eqnarray}
where $C$ is the constant equal to $8\pi/3$,
$a=a(\tau)$ and $\Phi=\Phi(\tau)$ are the scale parameter and the inflaton, respectively, and $\sigma$ stands
for the brane tension. If the brane tension tends infinity (physically, if $V_{char}/\sigma\ll 1$,
where $V_{char}$ is a characteristic value of $V$ along the solutions $\Phi(\tau)$)
one obtains the standard equations for the CdL instantons
in the Einstein's theory of relativity.
The solution of the
system (\ref{eebrane}) is called the CdL instanton if the (Euclidean) action
\begin{eqnarray}  \label{action}
& &
S=2\pi^2\int_0^{\tau_f}{\rm d}\tau
\left[a^3\left(\frac{1}{2}(\Phi')^2+V\right)+ \right. \nonumber \\
& &
\left. \frac{a^3}{2\sigma}\left(\frac{1}{2}(\Phi')^2+V\right)^2+
\frac{1}{C}\left( a^2a''+a(a')^2-a\right)\right]
\end{eqnarray}
is finite. The finiteness of the action is guaranteed by the following boundary conditions
\be \label{bc}
a(0)=0,a'(0)=1, \Phi'(0)=\Phi'(\tau_f)=0  ,
\ee
where $\tau_f>0$ is to be determined from $a(\tau_f)=0$. The action (\ref{action}) of the CdL instanton is
considerably simplified by using the equations of motion (\ref{eebrane}), \cite{jabv}
\be \label{action1}
S=-\frac{4\pi^2}{3C}\int_0^{\tau_f}a{\rm d}\tau-
\frac{\pi^2}{\sigma}\int_0^{\tau_f}a^3(\Phi')^4{\rm d}\tau .
\ee
The existence and the properties of a solution to the system (\ref{eebrane}), (\ref{bc}) is a very interesting question,
\cite{js}, \cite{tanaka}, \cite{weinberg}, \cite{jabv}, also in the limit $\sigma\to\infty$. For the definiteness, let
us denote by $\Phi_M$ the value of the inflaton where $V$ reaches the local maximum (the top) and
$V_M\equiv V(\Phi_M)$ is the related energy density of the scalar field.
There is always
the trivial solution called HM instanton
\begin{align}  \label{hminst}
&\Phi=\Phi_M,\ a= \HBM^{-1}\sin\left( \HBM\tau\right) , \quad \tau\in[0,\pi/\HBM]& \nonumber \\
&\HBM^2=\frac{8\pi}{3}\left( V_M+\frac{V_M^2}{2\sigma}\right)=H_M^2\left( 1+\frac{V_M}{2\sigma}\right),&
\end{align}
with the action \cite{jabv}
\be \label{hmaction}
S_{HM}=-\frac{\pi}{\HBM^2}=-\frac{3}{8V_M}\frac{1}{1+\frac{V_M}{2\sigma}},
\ee
where $H_M=(CV_M)^{1/2}$ is the Hubble parameter of the de Sitter universe filled by the energy density $V_M$ and
above defined $\HBM$ is the analogue of this quantity on considered brane. In the work of reference \cite{jabv} it
is shown that for
\be \label{critvalues}
\hat{\xi}\equiv \left. -\frac{1}{\HBM^2}\frac{{\rm d}^2V(\Phi)}{{\rm d}\Phi^2}\right|_{\Phi=\Phi_M}
\to l(l+3), \quad l=1,2,3,\dots
\ee
there is the solution of the CdL instanton equations such that the function $\Phi(\tau)$ crosses the top of the
barrier just $l$times - the $l$th order CdL instanton. (An analogical situation is also in the standard Einstein's
theory of relativity, with only change $\HBM\mapsto H_M$, \cite{tanaka}, \cite{vj1}, \cite{vj2} and \cite{weinberg}).
In the case $l=1$ the situation has been studied extensively in \cite{jabv}, where the approximative formulas for
the first order CdL instanton and its action have been found in the mentioned limit (\ref{critvalues}). Let us
recall the main results briefly.
If we introduce the notation $y=\Phi-\Phi_M$, then in the limit (\ref{critvalues}) we have
the CdL instanton that is close to the HM instanton (the limit CdL instanton) and in the lowest order we can write
$y=k\cos(\HBM \tau)$ and the
constant $k$ (the inflaton amplitude) can be determined perturbatively. The result is \cite{jabv}
\begin{eqnarray}  \label{k2}
& &
k^2=-\frac{7\left(4-\hat{\xi}\right)}
{16C+\frac{1}{24}\hat{\eta}^2+\frac{1}{2}\hat{\zeta}+\frac{1}{32}\frac{V_M}{\sigma}\left( 435-69C\right)}\equiv \nonumber
\\
& &
-\frac{7\left(4-\hat{\xi}\right)}{\mathfrak{D}},
\end{eqnarray}
where we have introduced the following quantities describing the properties of the potential $V$ at its top:
\be \label{hateta}
\hat{\eta}=\left. \frac{1}{\HBM^2}\frac{{\rm d}^3V(\Phi)}{{\rm d}\Phi^3}\right|_{\Phi=\Phi_M}, \
\hat{\zeta}=\left. \frac{1}{\HBM^2}\frac{{\rm d}^4V(\Phi)}{{\rm d}\Phi^4}\right|_{\Phi=\Phi_M} .
\ee
And the action of this instanton is given by the formula, \cite{jabv}
\begin{eqnarray} \label{scdl1}
& &
S_{CdL}=S_{HM}-\frac{4\pi^2}{3C}\frac{4C-5}{3}\frac{1}{\HBM^2}\frac{V_M}{\sigma}k^2= \nonumber \\
& &
S_{HM}+\frac{4\pi^2}{3C}\frac{4C-5}{3}\frac{1}{\HBM^2}\frac{V_M}{\sigma}\frac{7\left(4-\hat{\xi}\right)}{\mathfrak{D}} .
\end{eqnarray}
There are two possibilities: if ${\mathfrak D}>0$ then the limit CdL instanton exists for $\hat{\xi}>4$ and if
${\mathfrak D}<0$ then the limit CdL instanton exists for $\hat{\xi}<4$. Anyway, the action of such a limit CdL instanton
is less than the action of related HM instanton and therefore the CdL instanton mediates the vacuum decay.
The equations (\ref{k2}) and (\ref{scdl1}) can be
compared with their counterpart in the general theory of relativity, \cite{vj1},\cite{vj2}.
The expression (\ref{k2}) for the inflaton
amplitude approaches the general-relativistic value \cite{vj2} as $V_M/\sigma\to 0$.

\section{The limit CdL instanton of the second order}

In order to investigate the limit CdL instanton of the second order we introduce new independently variable
\bdis
x=\HBM \tau
\edis
and the expansions of the relevant quantities into the perturbative series in powers of the parameter $k$
\begin{align} \label{kexp}
&y(x)=\sum_n k^n u_n(x)& &a(x)=\HBM^{-1}\sum_{n} k^n v_n(x)& \nonumber \\
&\hat{\xi}=10+\sum_{n} k^n \Delta_n& &x_f=\pi+\sum_{n} k^n x_f^{(n)} . &
\end{align}
Inserting these series into the equations of motion (\ref{eebrane}) we obtain the infinite system of
linear equations for the functions $u_n$ and $v_n$
\begin{align*}
&u_n''(x)+3\cot(x)u_n'(x)+10u_n(x)={\mathcal U}_n(x)& \\
&v_n''(x)+v_n(x)={\mathcal V}_n(x)\sin(x),&
\end{align*}
where the source terms proportional to ${\mathcal U}_n$ and ${\mathcal V}_n$ are to be computed
order-by-order using (\ref{kexp}) and the Taylor expansion of the potential $V$ around $\Phi_M$
\bdis
V=\HBM^2\left[
\frac{V_M}{\HBM^2}-\frac{1}{2}\hat{\xi}y^2+\frac{1}{6}\hat{\eta}y^3+\frac{1}{24}\hat{\zeta}y^4+\dots\right] .
\edis
In the lowest order in $k$ we obtain easily
\begin{align} \label{v0u0v1}
&v_0=\sin(x)& &u_0=0& &v_1=0.&
\end{align}
The equation for $u_1$ is
\bdis
u_1''(x)+3\cot(x)u_1'(x)+10u_1(x)=0 .
\edis
The solution must obey the boundary conditions (\ref{bc}). In order to interpret $k$ as the amplitude of
the limit instanton in the $\Phi$-direction we can request that $|u_1(0)|(=|u_1(\pi)|)=1$. Then we have
\be \label{u1}
u_1(x)=\frac{1}{4}\left( 5\cos^2(x)-1\right) .
\ee
One straightforwardly obtains also the equation for $v_2$
\bdis
v_2''+v_2=-\left[ E (u'_1)^2-Fu_1^2\right]\sin(x),
\edis
with
\bdis
E=C+\frac{5-C}{4}\frac{V_M}{\sigma},\ F=5C\left( 1+\frac{V_M}{\sigma}\right) .
\edis
The boundary conditions (\ref{bc}) require that $v(0)=v'(0)=0$, therefore
\begin{eqnarray} \label{v2}
& &
v_2(x)=\frac{1}{6144}\left[ 24(100E-13F)x\cos(x)+\right. \nonumber \\ & &  \left. (-2800E+752F)\sin(x)+ \right. \nonumber \\
& &
\left.
15(20E-7F)\sin(3x)-25(4E+F)\sin(5x)\right] .
\end{eqnarray}
Having this result we can compute the shift of the right-end point $x_f$ in the second order in $k$. The result
reads
\be \label{xf2}
x_f^{(2)}=\frac{\pi}{64}\left[ -25E+\frac{13}{4}F\right]=\frac{5\pi}{256}
\left[ -7C+(-25+18C)\frac{V_M}{\sigma}\right] .
\ee
The explicit formula for the inflaton amplitude can be obtained from the equation for $u_2$ because $\Delta_1$ enters
this equation nontrivially. In fact, this equation reads
\begin{eqnarray} \label{u2eqs}
& &
u_2''(x)+3\cot(x)u_2'(x)+10u_2(x)= \nonumber \\
& &
-\frac{\Delta_1}{4}(5\cos^2(x)-1)+\frac{\hat{\eta}}{32}(5\cos^2(x)-1)^2 ,
\end{eqnarray}
or, with the help of more appropriate independently variable $z=\cos(x)$, (\ref{u2eqs}) has the
form of hypergeometric equation:
\begin{eqnarray*}
& &
(1-z^2)\frac{{\rm d}^2u_2(z)}{{\rm d}z^2}-4z\frac{{\rm d}u_2(z)}{{\rm d}z}+10u_2(z)= \\
& &
-\frac{\Delta_1}{4}(5z^2-1)+\frac{\hat{\eta}}{32}(5z^2-1)^2 .
\end{eqnarray*}
This equation has only polynomial solution that is bounded at $z=\pm 1$, namely
\bdis
u_2=\alpha +\beta z^4
\edis
where the constants $\alpha,\beta$ and the shift of the effective curvature of the potential $\Delta_1$ must obey
equations
\begin{align*}
&-18\beta=\frac{25}{32}\hat{\eta}&
&12\beta=-\frac{5}{4}\Delta_1-\frac{5}{16}\hat{\eta}&
&10\alpha=\frac{1}{4}\Delta_1+\frac{1}{32}\hat{\eta} .&
\end{align*}
One obtains easily the solution:
\begin{align*}
&\alpha=\frac{7}{960}\hat{\eta}&
&\beta=-\frac{25}{576}\hat{\eta}&
&\Delta_1=\frac{1}{6}\hat{\eta} . &
\end{align*}
Now, we can conclude that the the function $u_2$ is given by the formula
\be \label{u2}
u_2(x)=\frac{\hat{\eta}}{192}\left( \frac{7}{5}-\frac{25}{3}\cos^4(x)\right)
\ee
and the inflaton amplitude is given by
\be \label{k1l2}
\hat{\xi}=10+k\Delta_1 \quad \Rightarrow \quad k=\frac{6}{\hat{\eta}}\left( \hat{\xi}-10\right) .
\ee
The result (\ref{k1l2}) is not affected by the presence of the term(s) in the instanton equations
(\ref{eebrane}) proportional to the fraction $V_M/\sigma$. To compare this with the situation in the
Einstein's theory of relativity see \cite{vj2}, \cite{socdlja}. The formula (\ref{k1l2}) is singular for
$\hat{\eta}=0$. If this situation takes place, we must go back to the equation (\ref{u2eqs}) from which we see that the
requirement of the absence of resonance gives us $\Delta_1=0$. This means that $k$ remains
undetermined in this case and one should continue the computation to higher order in $k$. \\
We will finish this section with the computation of the difference between the actions of the above studied
second order CdL instanton and the related HM instanton. Both actions are given by the formula (\ref{action1}). In the
lowest non-zero order the mentioned difference of the actions is
\begin{eqnarray}
& &
\delta S\equiv S_{lim}^{(l=2)}-S_{HM}=-\frac{4\pi^2}{3C}k^2\int_0^{\tau_f}\HBM^{-1} v_2(\HBM \tau){\rm d}\tau= \nonumber \\
& &
-\frac{4\pi^2}{3C\HBM^2}k^2\int_0^\pi v_2(x){\rm d}x=
-\frac{4\pi^2}{3C\HBM^2}k^2\frac{25}{12}(C-1)\frac{V_M}{\sigma}= \nonumber \\
& &
-\frac{100\pi^2}{C\HBM^2}(C-1)\left(\frac{\hat{\xi}-10}{\hat{\eta}}\right)^2\frac{V_M}{\sigma} .
\end{eqnarray}
The proportionality of $\delta S$ to $V_M/\sigma$ was expected since in the Einstein's relativity theory
the analogical quantity is equal to zero, \cite{vj2} and \cite{socdlja}, in the second order in $k$.
$\delta S$ diverges as $\hat{\eta}\to 0$ - this is the same situation as within Einstein's theory of relativity.

\section{Conclusion}

We have derived explicit formula for the second order CdL instanton in the situation when the effective curvature of
the inflaton potential at its top is close to the critical value $10$. We have shown that the shape of the instanton
in our brane-world model is only small deformation of the shape of corresponding CdL instanton in the Einstein theory
of relativity. This deformation can be described by two changes in the parameters describing the function
$V=V(\Phi)$ close its local maximum: $\xi\mapsto\hat{\xi}$ and $\eta\mapsto\hat{\eta}$, where the {\it hat}-quantities
are affected by the dimensionless fraction $V_M/\sigma$ (as introduced in (\ref{hminst},\ref{critvalues},\ref{hateta}))
that describes the effect of the brane tension. We were also able to derive an explicit expression for the action of the
instanton and we have shown that in our brane-world model the difference between the action of our instanton and
related HM instanton appear in lower order in the $k$-expansion with respect to the situation in the general theory of
relativity.

\section*{Acknowledgments}
This work was supported by the Slovak Scienfific Agency, project VEGA 1/3042/06.

\end{document}